\documentclass[11pt, epsfig]{article}

\usepackage{amsmath}

\usepackage[hmargin=2.5cm,vmargin=2.5cm]{geometry}
\usepackage{graphicx}
\begin{document}
\title{Consistent thermodynamics for spin echoes}
\author {Charis Anastopoulos\footnote{anastop@physics.upatras.gr} \\
 {\small Department of Physics, University of Patras, 26500 Greece} \\ \\
 and \\ \\ Ntina Savvidou\footnote{ntina@imperial.ac.uk} \\
  {\small  Theoretical Physics Group, Imperial College, SW7 2BZ,
London, UK} }
\maketitle

\begin{abstract}
Spin-echo experiments are often said to constitute an instant of anti-thermodynamic behavior in a concrete physical system that violates the second law of thermodynamics. We argue that a proper thermodynamic treatment of the effect should take into account the correlations between the spin and translational degrees of freedom of the molecules. To this end, we construct an entropy functional using Boltzmann macrostates that incorporates both spin and translational degrees of freedom. With this definition there is nothing special in the thermodynamics of spin echoes: dephasing corresponds to Hamiltonian evolution and leaves the entropy unchanged; dissipation increases the entropy. In particular, there is no phase of entropy decrease in the echo. We also discuss the definition of macrostates from the underlying quantum theory and we show that the  decay of net magnetization provides a faithful measure of entropy change.
\end{abstract}

\section{Introduction}
The spin echo effect \cite{Hahn} has received significant attention
in relation to the foundations of statistical mechanics \cite{Roth, Blatt,
 Prig, BH, DD85, SK, Lands, RR, Lav, ZL}.
 It arguably
 provides a (partial) physical realization of  Loschmidt's velocity inversion paradox \cite{Losch}, albeit  in the
context of nuclear magnetism rather than of gases. For this reason, it is often stated that the spin-echo effect constitutes an
instance of a physical system manifesting anti-thermodynamic behavior. More complex echo phenomena, for example, \cite{Pines, SZ, PL}, provide a fuller realization of the Loschmidt inversion i.e., they involve the full many-body interactions. Moreover, echo phenomena are a testing ground for ideas on the origin of the irreversibility in macroscopic and mesoscopic systems. However, the issue of providing a quantitatively precise thermodynamic description arises even in the simplest case of Hahn echoes \cite{Hahn}.

The importance of the echo effects originates from the fact that, after the inversion of spins by the action of  external pulses, a macroscopic system seems to evolve spontaneously from a disordered state into an ordered one. This would appear to contradict the non-equilibrium version of the second law of thermodynamics. Many researchers   express the opinion that this contradiction is only apparent;  the evolution is truly irreversible and the   decay of magnetization provides a measure of entropy change during an echo. However, it is difficult make this statement quantitatively precise. The usual notions of entropy (coarse-grained Gibbs entropy or Boltzmann entropy), when applied to a spin system, show a decrease of entropy after spin inversion \cite{RR, Lav}. Moreover, they bear no relation to the echo decay that is thought to provide a measure of irreversibility.
In this paper, we emphasize the necessity of a precise thermodynamic description, where the entropy functional that always remains a non-decreasing function of time during an echo experiment.

To this end, we follow Boltzmann's definition of entropy and we specify macrostates relevant to the system. We argue that a consistent definition of macrostates cannot separate the spin from the translational degrees of freedom, because they become non-trivially correlated in the course of  a spin-echo experiment. Hence, by adapting Boltzmann's coarse-graining we define an entropy functional that provides a thermodynamic description of spin-echo experiments, with no violation of the second law.
Furthermore, using simple quantum open system dynamics for the description of relaxation, we show that the entropy increase is indeed a monotonically decreasing function of magnetization decay in an echo.

For the description of the Hahn spin echoes it suffices, as a first approximation,  to treat the thermodynamic system as an assembly of
non-interacting microsystems. Interactions are essential for the understanding of irreversibility, but they do not affect the issue whether there is a phase of
decreasing entropy after spin inversion \cite{Lav}.
We, therefore, consider
 an assembly of {\em classical }
magnetic dipoles precessing in a magnetic field. The assembly consists of N particles with magnetic
moments ${\bf m}_i = m  ( \cos \theta \sin \phi, \cos \theta
\cos \phi, \sin \theta)$, in terms of the spherical coordinates $\theta$ and $\phi$; $m $ is  constant equal to $gs$, where $g$ is the gyromagnetic ratio and $s$ the magnitude of the particle's classical spin vector.

 The total magnetization of the system is ${\bf
M} = \sum_i {\bf m}_i$. If a constant magnetic field $B$
is applied along the $z$ axis, the equilibrium configuration at
temperature $\beta^{-1}$ consists of all dipoles oriented along the
direction of the field, provided that $\beta m B
>> 1$. A $\frac{\pi}{2}$-pulse is then applied on the system rotating the magnetic
moments by $\frac{\pi}{2}$ so that they  become oriented along the
$x$-axis. At this moment ($t = 0$), the magnetic moment of each
dipole equals ${\bf m}_i = m (1, 0, 0)$ and a strong
magnetization ${\bf M} (0) = N m(1, 0, 0) $ along the $x$ axis is measured.

The dipoles then precess around the $z$ axis according to the
equation
\begin{eqnarray}
\dot{{\bf m}}_i =  g {\bf m}_i \wedge {\bf B}_i, \label{dipole}
\end{eqnarray}
 where $g$ is the gyromagnetic ratio. Each dipole precesses with a different   value ${\bf B}_i$ of the magnetic field, reflecting the fact that the magnetic field
  ${\bf B}$ is not homogeneous within the material.

By solving Eq. (\ref{dipole}) we obtain
\begin{eqnarray}
{\bf m}(t) = m (\cos\omega_i t, \sin \omega_i t, 0),
\end{eqnarray}
where $\omega_i = g B_i$ is the angular frequency of precession for
the dipole $i$.  The $x$ component of the
magnetization is then $ M_x(t) = m \sum_i \cos \omega_i t$. Assuming that the  angular frequencies $\omega_i$ are randomly distributed in an
interval $[\omega_{min}, \omega_{max}]$, $M_x(t)$
rapidly becomes vanishingly small because of  dephasing.

At
 $t = \tau$, a $\pi$-pulse is applied on the system, so that the dipoles'
 configuration is transformed as
\begin{eqnarray}
m (\cos \omega_i \tau, \sin \omega_it, 0) \xrightarrow{} m (\cos
\omega_i \tau, - \sin \omega_i \tau, 0),
\end{eqnarray}
   i.e., the pulse inverts the dipoles' orientation on the $x-y$ plane. After
     inversion the dipoles precess freely. Hence,
\begin{eqnarray}
{\bf m}_i(t) =  m (\cos(\omega_i t - 2 \omega_i \tau), \sin
(\omega_i t - 2 \omega_i \tau), 0). \label{mm}
\end{eqnarray}

At $t = 2 \tau$, Eq. (\ref{mm}) predicts that ${\bf m}_i = m (1, 0
,0)$, i.e.,  there is strong magnetization in the $x$ direction, the
same as   at time $t = 0$.
 Apparently, the system starts from an `ordered' state, it evolves
 into a `disordered' one at time $t = \tau$, but after the inversion
 it evolves back to the initial `ordered' state. Hence, it seems as though
 the system evolves from a
  disordered into an ordered state, without any external action during the time interval $[\tau, 2 \tau]$.

The quantification of the latter statement requires the definition of a
{\em non-equilibrium} entropy function for the system. If the
entropy is defined in terms of the dipole degrees of freedom
alone, then the inevitable conclusion is that the disordered state at $t
= \tau$ is of higher entropy than the ordered state at $t = 2 \tau$---see Refs. \cite{RR, Lav} and also Sec. 2. It
follows that during the time interval $[\tau, 2 \tau]$ the system
evolves spontaneously to states of lower entropy. This is  a manifestation of  anti-thermodynamic behavior.

In spin-echo
experiments, the cause of dephasing  is the spatial inhomogeneity of the magnetic field. In statistical mechanics, an external magnetic field is treated as an  constraint external to the thermodynamical system. In particular, in spin-echoes  the magnetic field constraint distinguishes the dipoles by their position. Position is not an abstract label, it is as much a physical degree of freedom  as spin is.  The treatment the dipole degrees of freedom in
isolation presupposes: (i) a decoupling between translation and dipole
degrees of freedom, and (ii)  that any correlations
between them are insignificant. Condition (i) is not satisfied, however the translational degrees of freedom can be considered---to a good approximation---as a bath inducing dissipation and noise on the dipoles' evolution. Condition (ii) is more problematic in the sense  the inhomogeneity of the field creates
correlations between position and dipole degrees of freedom. The isolation of the dipole degrees of freedom is a drastic simplification because it removes all information about such correlations from the entropy function.

 The above argument strongly suggests that a consistent thermodynamic description of the
 spin-echo effect should involve a state space that also incorporates translational degrees of freedom. We demonstrate than in this case, we can define a Boltzmann entropy that   accounts for the correlations between  magnetic moments and position. Hence, we conclude that in a spin-echo experiment, the information of the ordered initial state is transferred into information about non-trivial correlations \cite{Prig}. After the application of the  $\pi$-pulse, this information in correlations is again transferred to information about spin order. Hence,  a phase
of decreasing entropy never appears.

The structure of this paper is the following. In Sec. 2, we present  the definition of an entropy function in terms of Boltzmann macrostates that include translational degrees of freedom; this definition leads to a description of spin echoes with no apparent violation of the second law of thermodynamics. In Sec. 3, we   define this version of Boltzmann entropy function in quantum theory. In Sec. 4, we show that the Boltzmann entropy is an increasing function of time when relaxation effects are also taken into account. Finally in Sec. 5, we summarize and discuss our results.

\section{Boltzmann entropy and macrostates for a spin system}

In this section, we define an entropy function for spin systems that provides a consistent thermodynamic description of the spin-echo experiments. As explained in the Introduction, the key idea is that the entropy should also incorporate the correlations between spin and translational degrees of freedom, for the molecules.

First, we examine the evolution
 of Boltzmann entropy for the model of Sec. 1, where the translational degrees of
freedom are ignored. The assembly of dipoles is labeled by the
abstract index $i$. Since the dipoles are independent, their statistical behavior can be described by a distribution function $f(\theta, \phi)$, defined on the two-sphere $S^2$ of directions in space.

  Assuming that at $t = 0$ all dipoles are
oriented along the $x$-axis, then the distribution function for a single dipole approximates a delta function $\delta_{S^2}( \theta -
\frac{\pi}{2}, \phi)$ on
the sphere $S^2$. The dipole's precession preserves $\theta$, hence, the
delta-function evolves as $\delta (\theta - \frac{\pi}{2}) \sum_{n =
-\infty}^{\infty} e^{i n (\phi - \omega_i t)}$. Since the dipoles
are independent, the system's macrostate  is
described by the averaged density
\begin{eqnarray}
f(\theta, \phi) = \frac{1}{N} \delta( \theta - \frac{\pi}{2}) \sum_i
\sum_n e^{i n (\phi - \omega_i t)}.
\end{eqnarray}

The detailed evolution of $f$ depends on the distribution of frequencies.
We assume  a Gaussian distribution
around a mean frequency $\bar{\omega}$---corresponding to the mean
magnetic field---with a deviation $\sigma_{\omega} << \bar{\omega}$. Then,
\begin{eqnarray}
f(\theta, \phi, t) = \delta( \theta - \frac{\pi}{2})
\frac{1}{\sqrt{2 \pi
 \sigma_{\omega}^2}} \sum_n \int d \omega e^{- \frac{(\omega -
 \bar{\omega})^2}{2 \sigma_{\omega}^2} + i  n (\phi - \omega_i t)} =
\delta( \theta - \frac{\pi}{2}) \sum_n e^{i n (\phi - \bar{\omega}
t) - \frac{n^2\sigma_{\omega}^2t^2}{2}}. \label{ff}
\end{eqnarray}
Eq. (\ref{ff}) corresponds to a  diffusive evolution equation
\begin{eqnarray}
\frac{\partial f}{\partial t} = - \bar{\omega} \frac{\partial
f}{\partial \phi} + \sigma^2_{\omega} t \frac{\partial^2f}{\partial \phi^2},
\end{eqnarray}
which is entropy-increasing
 \begin{eqnarray}
 \dot{S}_B = \sigma_{\omega}^2 t \int d^2 s \frac{1}{f}
\left( \frac{\partial f}{\partial \phi} \right)^2 \geq 0.
\end{eqnarray}

  For the case of dipoles that are distinguished by  variables external to the system, i.e., the label $i$, we  employ an averaging procedure that leads to an entropy-increasing evolution equation for the probability density. It follows the  entropy  at time $t = \tau$, when the pulse is applied, is larger than the entropy at time $t=0$. Hence, when the system retraces its past evolution after $t = \tau$, entropy decreases.

  As mentioned earlier, the actual  distinction of the dipoles arises from the inhomogeneity of the field in space, i.e., from interactions involving translational degrees of freedom of the molecules. Therefore, the translational degrees of freedom should also be incorporated into the definition of the macrostates  describing the system.

To this end, we recall Boltzmann's prescription for the macrostates and entropy of rare gases.   For a rare gas
of $N$ particles, microstates correspond to points of the state
space $\Gamma = {\bf R}^{6N}$. Macrostates correspond to
distribution functions $f({\bf x}, {\bf p})$ on the state space of a
single particle $\Gamma_1 = {\bf R}^6$.

 The Boltzmann macrostates
are constructed as follows \cite{Boltz}: one splits the
$\Gamma_1$-space into cells $C_a$ of volume $(\Delta x)^3 (\Delta
p)^3 >> 1$. To each macrostate $\xi = ({\bf x}_1, {\bf p}_1; {\bf x_2}, {\bf p}_2; \ldots;
{\bf x}_N, {\bf p}_N) \in \Gamma$ we
assign the sequence of numbers $n(C_a)$
corresponding to the number of particles such that  $({\bf x}, {\bf
p}) \in C_a$. The sequence $n(C_a)$ fully specifies a macrostate for
the system. At the continuum limit, the sequence $n(C_a)$ defines a probability distribution $f({\bf x}, {\bf p})$ on $\Gamma_1$. In
what follows, it is convenient to  choose  $f({\bf x}, {\bf p})$ normalized to
unity.

 The Boltzmann
entropy $S_B$ is the logarithm of the number of microstates in each
macrostate, hence for any sequence $n(C_a)$, $S_B \sim \sum_a \ln
n(C_a)$. At the continuum limit,
\begin{eqnarray}
S_B[f] = - N \int d^3x \, d^3p \; f({\bf x}, {\bf p}) \; \ln f({\bf
x}, {\bf p}) + N - N \ln N.
\end{eqnarray}

The key point  is that  Boltzmann's definition of macrostates  applies to any  system of $N$ particles, {\em even} when the particles  have degrees of freedom other than the translational ones. It suffices that the state space of the system can be expressed as
  $\Gamma =
\Gamma_1^N$. The dipole degrees of freedom corresponds to
 a classical
spin vector ${\bf s}$ of constant norm $s$, i.e., the corresponding
state space is the sphere $S^2$ with area equal to $2s$. The two-sphere is a
symplectic manifold with symplectic form
$\Omega = \frac{s}{2 \pi} \sin \theta d \theta \wedge d \phi$. Hence, the equations of motion for the dipole degrees of freedom are Hamiltonian.

It follows that the classical state space of a single particle with spin is $\Gamma_1 =
{\bf R}^6 \times S^2$. The  state space for $N$-particles with spin
$\Gamma = {\bf R}^{6N}\times (S^2)^N$ consists of points $\xi =
({\bf x}_1, {\bf p}_1, {\bf s}_1;   \ldots; {\bf x}_N,  {\bf p}_N,  {\bf s}_N)$.
Next, we apply Boltzmann's analysis in a straightforward way.
  At the continuum limit, a macrostate is described by a
distribution $f({\bf x}, {\bf p}, {\bf m})$ on $\mu$, normalized to
unity. The Boltzmann entropy  is then a functional of $f$
\begin{eqnarray}
S_B[f] = - N \int d^3x \, d^3p \, d^2 s \; f({\bf x}, {\bf p}, {\bf
s}) \; \ln f({\bf x}, {\bf p}, {\bf s}) + N - N \ln N,
\label{entropy}
\end{eqnarray}
where   $d^2s = \frac{s}{2 \pi} \sin \theta d \theta d \phi$.

 Next, we consider the dynamical evolution of the distribution  $f$ in accordance to  the simplified model of Sec. 1. We assume that the particle system is subjected to an external inhomogeneous magnetic field ${\bf
B}({\bf x}) = B({\bf x}) {\bf \hat{z}}$. Here, we  ignore spin-spin and spin-lattice interactions, as well as the action of the magnetic field
on the translational degrees of freedom. We also assume that the
translational degrees of freedom are initially in a state of thermal
equilibrium. Then,
  the distribution $f({\bf x}, {\bf p}, {\bf s})$ changes in time
  only because of the spin's precession around the magnetic field's axis.
  In terms of
  the spherical coordinates $\theta, \phi$,
  the time evolution law for this class of states, is $f({\bf x}, {\bf p}, \theta, \phi) \rightarrow f({\bf x}, {\bf p},
    \theta, \phi - g B({\bf x}))$, or, equivalently
  \begin{eqnarray}
   \frac{\partial f}{\partial t}= - g B({\bf x})  \frac{\partial f}{\partial
   \phi}. \label{eqmotion}
  \end{eqnarray}
The key difference of Eq. (\ref{eqmotion}) from Eq. (\ref{ff}) is that in Eq. (\ref{eqmotion}), ${\bf x}$ is a variable of $f$ and not an external label. Hence, the necessity to average over different values of $f$ in order to obtain a closed evolution equation does not arise.
Substituting Eq. (\ref{eqmotion}) into Eq. (\ref{entropy}) we obtain
\begin{eqnarray}
\dot{S}_B [f] =  \frac{N  s }{2 \pi}  \int d^3x \, d^3p \, \sin
\theta d \theta \;
 B({\bf x}) \; \left(\int d \phi \frac{\partial (f \ln f)}{\partial \phi}\right)  =
 0.
\end{eqnarray}
Therefore, we showed that the Boltzmann entropy remains constant, irrespective of whether
there is dephasing or rephasing of the transverse magnetic moments. Moreover, the action of the $\pi$-pulse does not change the entropy.
Hence, the Boltzmann entropy records no violation of the second law of
thermodynamics during the rephasing stage of a spin-echo experiment.
We therefore conclude that if
the particle positions are included into the definition of the
system's macrostates, the system's evolution is
reversible.

 While Eq. (\ref{eqmotion}) is reversible, it does not
coincide with Liouville's equation for a single dipole coupled to
the inhomogeneous magnetic field ${\bf B}({\bf x})$. The Liouville equation for the Hamiltonian  $H = -  {\bf B}({\bf x}) \cdot {\bf m} = -  m
B({\bf x}) \sin \theta$
involves an additional term $ - m {\bf \nabla}B({\bf x})\cdot
\frac{\partial f}{\partial {\bf p}}\sin \theta$, in the
right-hand-side of Eq. (\ref{eqmotion}).

We proceed to show that the additional term is   small, so that Eq.
(\ref{eqmotion}) well approximates the Liouville equation. Since we assume that the translational degrees of freedom are
in thermal equilibrium, the momentum dependence of the distribution
$f({\bf x}, {\bf p}, \theta, \phi)$ is of the
Maxwell-Boltzmann type $\sim e^{- \beta \frac{{\bf p}^2}{2M}}$,
where $M$ is the particles' mass. Denoting by $A(f)$ the right-hand-side term  of Eq. (\ref{eqmotion}), and by $C(f)$ the additional
term above, we find that $|C(f)/A(f)|$ is of the order of $
\frac{\beta \sqrt{\overline{E}/M}}{M }$, where $L$ is the
characteristic scale of inhomogeneities in the magnetic field and
$\overline{E}$ the mean kinetic energy of a particle. The latter is of the order  of $\beta^{-1}$,  hence,  $|C(f)/A(f)| \sim \sqrt{\beta/M} / L$. Therefore, the term $C(f)$ is
negligible, if the scale of the inhomogeneities of the magnetic
field is much larger than the thermal de Broglie wavelength
$\lambda_{deB} \sim \sqrt{\beta/M}$ of the particles.

\section{Quantum Boltzmann entropy for spin-echo systems}

The definition of the function $f({\bf x}, {\bf p}, {\bf s})$ that
describes the system's macrostates incorporates Boltzmann's coarse-graining
 on the classical state space. However, spin is
fundamentally a quantum variable, and unlike the translational
degrees of freedom it has discrete spectrum. The
question then arises, how $f({\bf x}, {\bf p}, {\bf s})$ can be
constructed in terms of the underlying quantum theory. The aim of this section is to provide one such construction for the distribution function. We also  discuss some subtle points regarding the physical meaning of the corresponding macrostates.

\subsection{The description of spin-echo macrostates in quantum theory}In  derivations of the Boltzmann equation for quantum gases,
the corresponding distribution function is usually defined in terms of the
single-particle reduced density matrix \cite{QBEW, QBE}. In
particular, let ${\cal H} = {\cal L}^2({\bf R}^3) \otimes {\bf C}^{2s+1}$
be the  Hilbert space for a single particle of spin $s =
\frac{n}{2}, n =   1, 2, \ldots,$. A system of $N$ particles is
described by vectors on the Hilbert space $(\otimes H)^N$. The density matrix $\hat{\rho}$ of the $N$-particle system
is totally symmetrized for bosonic particles and totally
antisymmetrized for fermionic particles. The single-particle reduced
density matrix on ${\cal H}$ is defined as $\hat{\rho}_1 = Tr_{(\otimes H)^{N-1}} \hat{\rho}$.

Given the single-particle density matrix $\hat{\rho_1}$, we
construct different versions of the functions $f({\bf x}, {\bf p},
{\bf s})$ according to the theory of quantum quasi-probability
distributions. In derivations of the quantum Boltzmann equation
for gases the Wigner function is usually employed \cite{QBEW},
mainly because it simplifies the calculations.
 A Wigner
function on the single-particle state space $\Gamma_1 = {\bf
R}^6\times S^2$ is indeed defined from the reduced density
matrix $\hat{\rho}_1$ \cite{qpd}.
  However, Wigner functions are not positive-valued in general,
therefore: (i) they do not have a physical interpretation in terms
of particle number as in Boltzmann's definition of macrostates, and
(ii) they cannot be used to define Boltzmann entropy according to
Eq. (\ref{entropy}). A coarser quasi-probability distribution that is
positive-definite should be used instead. In general,
such distributions are constructed from Positive-Operator-Valued
Measures (POVMs) \cite{QMT}. Any family of positive operators,
 $\hat{\Pi}  ({\bf x}, {\bf p}, {\bf
s})$ normalized to unity as
\begin{eqnarray}
\int d \mu({\bf x}, {\bf p}, {\bf s}) \hat{\Pi}  ({\bf x}, {\bf p},
{\bf s}) = 1,
\end{eqnarray}
for some invariant measure $d \mu$ on $\Gamma_1$, defines a
mathematically appropriate probability distribution

\begin{eqnarray}
f({\bf x}, {\bf p}, {\bf s}) = Tr [\hat{\rho}_1 \hat{\Pi}  ({\bf x},
{\bf p}, {\bf s})]. \label{povm}
\end{eqnarray}

The simplest case of such a POVM is obtained from the coherent
states $|{\bf x}, {\bf p}, {\bf s} \rangle $ on ${\cal H}$,  setting
$\hat{\Pi} ({\bf x}, {\bf p}, {\bf s}) = |{\bf x}, {\bf p}, {\bf s}
\rangle \langle {\bf x}, {\bf p}, {\bf s}| $, so that
\begin{eqnarray}
f({\bf x}, {\bf p}, {\bf s}) = \langle{\bf x}, {\bf p}, {\bf s}|
\hat{\rho}_1|{\bf x}, {\bf p}, {\bf s} \rangle. \label{fq}
\end{eqnarray}
The  density $f({\bf x}, {\bf p}, {\bf s})$ is the Husimi
distribution associated to the single-particle reduced density
matrix $\hat{\rho}_1$ through the coherent states $|{\bf x}, {\bf
p}, {\bf s} \rangle $.

 The coherent states $|{\bf x}, {\bf p}, {\bf s} \rangle$
are defined as the tensor product $| {\bf x}, {\bf p} \rangle
\otimes |{\bf s} \rangle$, where $| {\bf x}, {\bf p} \rangle $ are
the standard coherent states on ${\bf L}^2({\bf R}^3)$ and $|{\bf s}
\rangle := |\theta, \phi \rangle$ are the spin-coherent states on
${\bf C}^{2s+1}$ \cite{CS}
\begin{eqnarray}
|\theta, \phi \rangle = \sum_{m_s = -s}^s \left( \begin{array}{c} 2s
\\ s+m \end{array} \right) \cos^{s + m_s} \frac{\theta}{2} \sin^{s -
m_s} \frac{\theta}{2} e^{-i m_s \phi/2} | m_s \rangle. \label{coh}
\end{eqnarray}

In Eq. (\ref{coh}),  $|s, m_s \rangle$  are the eigenstates of the $\hat{S}_z$
generator in the $(2s+1)$-dimensional representation of
 $SU(2)$. The resolution of the unity for the spin
 coherent states is
 \begin{eqnarray}
(2 s + 1 )\int  \frac{\sin \theta d \theta d \phi}{4 \pi} |\theta,
\phi\rangle \langle \theta, \phi| = 1.
 \end{eqnarray}

Hence, the invariant measure $d^2s$ equals $(2 s +1) \frac{\sin \theta d \theta
d \phi}{4 \pi}$ and  the volume of the corresponding two-sphere is
$(2 s + 1)$.

Since the particles are assumed independent, then the single-particle
reduced density matrix evolves under the Hamiltonian $\hat{H} = - g
B({\bf \hat{x}}) \mu \hat{S_z}$, where $\mu$ is the particle's
magnetic moment. A spin coherent state $|\theta, \phi \rangle$ in an
external magnetic field along the $z$-direction evolves into
another coherent state with parameters following  the corresponding
classical equations of motion, i.e., $|\theta, \phi \rangle
\rightarrow |\theta, \phi + g B   t \rangle$. As in the classical case, the assumption that the
inhomogeneity scale of the magnetic field is much larger than   the thermal de Broglie wavelength of the particles suffices to
guarantee that the distribution $f({\bf x}, {\bf p}, {\bf s})$ evolves under Eq.
(\ref{eqmotion}).

\subsection{Interpretation of the quasi-classical description}
In Sec. 3.1  we showed a suitable definition for the function
$f({\bf x}, {\bf p}, {\bf s})$ from the underlying quantum
description. However, we must elaborate here on the
 adequacy of the description of a quantum system by a
classical variable.  We ignore the
translational degrees of freedom and focus on the spin ones, so that
we express the POVM (\ref{povm}) simply as $\hat{\Pi}({\bf s})$. The classical state
space for a single spin is the sphere $S^2$. If $C$ is a region of
$S^2$ then we define $\hat{\Pi}_C = \int_C d^2 s \hat{\Pi}({\bf s})$.
The positive number $Tr (\hat{\rho_1} \hat{\Pi}_C)$ is interpreted as
an approximate probability that the  spin vector lies in the region $C$. Thus the assignment $C\rightarrow \hat{\Pi}_C$ defines an approximate correspondence between operators and state-space regions.
A necessary (but not sufficient) condition for this
correspondence to be meaningful is
that the volume $[C] = \int_C d^2 s$ of the region $C$
is much larger than unity \cite{Omn8894}. A
correspondence between quantum and classical observables exists only for
coarse-grained phase space regions of volume much larger than
$\hbar$.  The  area of the spin two-sphere is $ \hbar (2 s +
1)$, so  a proper correspondence between
operators and regions is only possible for values $s >> 1$.

Therefore, it seems that it is not possible to define a
classical description for particles with low values of spin---in
particular  spin $s = \frac{1}{2}$. The answer to this problem lies
in the remark that $\hat{\rho}_1$ is not the
density matrix of a single particle but the reduced density matrix
in a system of $N$ particles. We express the probabilities obtained from $\hat{\rho}_1$ in terms of the density matrix
$\hat{\rho}$ of the total system, and we obtain that $Tr (\hat{\rho}_1
\hat{\Pi}_C) = Tr (\hat{\rho} \hat{P}_C)$.
\begin{eqnarray}
\hat{ P}_C = \hat{\Pi}_C \otimes 1 \otimes \ldots \otimes 1 + 1
\otimes \hat{\Pi}_C  \otimes \ldots \otimes 1 + \ldots + 1 \otimes 1
\otimes \ldots \otimes \hat{\Pi}_C, \label{qcg}
\end{eqnarray}
is a positive operator corresponding to the proposition that  ``the spin of at least one particle takes values in the region $C$''.
$Tr \hat{\rho} \hat{ P}_C$ is therefore the corresponding
probability. The positive operator $\hat{P}_C$
corresponds to the region $ {\cal C} = (C \times S^2 \times \ldots \times S^2)
\cup (S^2 \times C \times \ldots \times S^2) \cup \ldots \cup (S^2 \times S^2
\times \ldots \times C)$ within the classical state space $(S^2)^N$ for the
$N$ spins. The volume of ${\cal C}$ is therefore
\begin{eqnarray}
[{\cal C} ]=   Tr \hat{P}_C = (N-1) (2 s+1) [C].
\end{eqnarray}

Hence, for systems with a large number of particles, the volume of
the region $[{\cal C}]$ can be significantly larger than unity, and the approximate correspondence between positive operators and
state space regions is meaningful.

\subsection{Quantum definition of spin-echo Boltzmann entropy.}

The distribution $f({\bf x}, {\bf p}, {\bf s})$ is
positive-valued. Therefore,  we can define the Boltzmann entropy as in Eq.
(\ref{entropy})
\begin{eqnarray}
S_B = - N \int d^3 x d^3 p \frac{(2s+1) \sin \theta d \theta d
\phi}{4 \pi} f({\bf x}, {\bf p}, \theta, \phi) \ln  f({\bf x}, {\bf
p}, \theta, \phi) + N - N \ln N. \label{sb}
\end{eqnarray}

We note that the expression
\begin{eqnarray}
S_W[\hat{\rho}_1] = - \int d^3 x d^3 p \frac{(2s+1) \sin
\theta d \theta d \phi}{4 \pi} f({\bf x}, {\bf p}, \theta, \phi) \ln
f({\bf x}, {\bf p}, \theta, \phi)
\end{eqnarray}
 is the Wehrl entropy,
associated to the coherent
states $|{\bf x}, {\bf p}, \theta, \phi \rangle$ \cite{Wehrl}. Hence,
\begin{eqnarray}
S_B = N S_W[\hat{\rho}_1] + N - N \ln N. \label{We}
\end{eqnarray}

The description of macrostates in terms of the single-particle reduced density matrix $\hat{\rho}_1$ does not  represent accurately the Boltzmann definition of macrostates, as described in Sec. 2. The coarse-graining corresponding to  $\hat{\rho}_1$ is determined by the projectors Eq. (\ref{qcg}), which represent the statement
``at least one particle is characterized by values
of the observables that correspond to the region $C \in \Gamma_1$''. In contrast, Boltzmann macrostates are defined in terms of the {\em number} of particles in $C$ at a moment of time. The equivalence between these coarse-grainings requires the assumption that statistical fluctuations in the number of particles in $C$ and  particle correlations are negligible; then, the probabilities $Tr(\hat{\rho}_1 \hat{P}_C)$ are indeed proportional to the number of particles in $C$.
The latter distinction is not specific to the construction presented here, but it reflects  the difference between the original derivation of Boltzmann's equation and the alternative derivation through the truncation
of the Bogoliubov-Born-Green-Kirkwood-Yvon hierarchy of correlation functions. It is in fact the reason why the two approaches employ {\em different} physical conditions for the domain of validity of Boltzmann's equation.

We believe that Boltzmann's coarse-graining is conceptually more satisfying, however in this paper we have chosen to work with the single-particle reduced density matrix. One reason is the significant technical difficulty in implementing Boltzmann's coarse-graining in quantum theory. In particular, an implementation of Boltzmann's coarse-graining from first principles requires a proof of decoherence (the corresponding variables behave quasi-classically), a property that is likely to require significant restriction to the initial states of the system---see Refs. \cite{Omn8894, Hal}. More importantly, the difference above is not significant for the Hahn echoes studied here because the $\pi$-pulse inversion does not affect the part of the Hamiltonian that corresponds to the many-body interaction. Hence,  a detailed description of the generation of irreversibility through spin-spin interactions is not necessary.

It is important to note that  the difference between the two coarse-grainings is more pronounced in spin systems than in gases. In principle, it is possible that in some systems the different coarse-grainings lead to different predictions. The reason is that in Boltzmann's coarse-graining, there is no kinematical restriction on  the distribution function $f({\bf s})$ on $S^2$. On the other hand, the form of the distribution function $f({\bf s})$, corresponding to the single-particle reduced density matrix, is constrained by the value of the spin $s$. $f({\bf s})$ belongs to the subspaces of ${\cal L}^2(S^2)$ corresponding to $s$. For example, for spin $s = \frac{1}{2}$, the function $f(\theta, \phi) = \langle \theta, \phi|\hat{\rho}_1|\theta, \phi \rangle$ can only be of the form
\begin{eqnarray}
f(\theta, \phi) = \sin^2\frac{\theta}{2} + x \cos \theta + r \sin \theta \cos(\phi + \chi),
\end{eqnarray}
for some constants $x, r, \chi$, while there is no restriction in the form of $f$ when  defined  in terms of Boltzmann's coarse-graining. Hence, evolution equations obtained through Boltzmann coarse-graining may not have a representation in terms of the single-particle reduced density matrix of the system. This issue will be explored elsewhere, in relation to the Loschmidt echoes.

Given the single-particle density matrix $\hat{\rho}_1$, one also has the option of employing  the von Neumann entropy $S_{vN}(\hat{\rho}_1) = - N Tr(\hat{\rho}_1 \log \hat{\rho}_1)$.  In particular, the irreversible evolution equations considered in Sec. 4   lead to an increase of the von Neumann entropy. However, given the fact that the Boltzmann coarse-graining in a spin system could lead to a distribution function that does not correspond to the single-particle density matrix, we consider that the
 Boltzmann entropy Eq. (\ref{We}) is a better candidate for the non-equilibrium entropy of the system.

\section{Entropy increase in  spin echoes }

\subsection{Evolution equation for the distribution function}
The description of the spin system described in Sec. 2 ignored
the relaxation effects that characterize the evolution of nuclear
spins. The inclusion of such effects requires the derivation of effective equations for the evolution of the distribution function $f({\bf x}, {\bf p}, {\bf s})$.

There are two main sources of irreversibility in the evolution of
the spin system: interactions between spins (dipole-dipole coupling)
and interactions between the spin and the translational degrees of
freedom (spin-lattice interaction). The former processes are
responsible for the relaxation of the transverse components of the
magnetization. The latter processes are responsible for the approach
to thermal equilibrium. We assume that the translational
degrees of freedom are in a state of thermal equilibrium, so that they
essentially act as a thermal reservoir for the spin variables.

Irreversible evolution is often described by  a master
equation for the single-particle reduced density matrix
$\hat{\rho}_1$. To this end, one  invokes a random field approximation, i.e., the assumption that each dipole evolves separately in a random magnetic field
$\hat{b}(t)$, generated by the other particles. Assuming that the autocorrelation
time of the random field is negligible, one employs the Markov
approximation in order to obtain a master equation of the Lindblad type
\cite{Lind}. The simplest such master equation is \cite{Open}
\begin{eqnarray}
\frac{\partial \hat{\rho}}{\partial t} = - i g   B [\hat{S}_z,
\hat{\rho}] - \frac{\Gamma_T}{2} \alpha \left([\hat{S}_-,
[\hat{S}_+, \hat{\rho}]] + [\hat{S}_+ [\hat{S}_-,
\hat{\rho}]]\right) \nonumber \\
- \frac{\Gamma_T}{2} \left( [\hat{S}_-, \hat{\rho}\hat{S}_+] -
[\hat{S}_+,
  \hat{S}_- \hat{\rho}]\right) -  \frac{\Gamma_L}{2} [\hat{S_z}, [\hat{S}_z,
\hat{\rho}]], \label{eq1}
\end{eqnarray}
where $\Gamma_T$ and $\Gamma_L$ are phenomenological
dissipation constants, corresponding respectively to the transverse and
longitudinal components of the random magnetic field.
  The parameter $\alpha$ is determined by the
requirement that the stationary solution to Eq. (\ref{eq1}) is a thermal
state at temperature $\beta^{-1}$
\begin{eqnarray}
\alpha = \frac{1}{e^{\beta g    B} - 1}. \label{alfa}
\end{eqnarray}

Eq. (\ref{eq1}) yields the phenomenological magnetic Bloch equations for the
macroscopic magnetization $M_i = N \langle \hat{S}_i\rangle$
\begin{eqnarray}
\dot{M}_z &=&   \Gamma_T \left[ (-\frac{1}{2}) - (2 \alpha + 1)
M_z\right] \\ \label{bloch1}
\dot{M}_{\pm} &=& \mp i g  B M_{\pm} - \left(\Gamma_T (\alpha
+ \frac{1}{2}) + \Gamma_L \right) M_{\pm}. \label{bloch2}
\end{eqnarray}
The Bloch equations are usually expressed in terms of the
spin-lattice relaxation time $T_1$   and the spin-spin relaxation
time $T_2$, which are identified from Eqs. (\ref{bloch1}-\ref{bloch2}) as

\begin{eqnarray}
T_1 = [ \Gamma_T (2 \alpha +1)]^{-1} \label{t1}\\
T_2 = [\Gamma_T (\alpha + \frac{1}{2}) + \Gamma_L ]^{-1} \label{t2}.
\end{eqnarray}

For the case $s = \frac{1}{2}$, we obtain the solutions to Eq. (\ref{eq1})
\begin{eqnarray}
\rho_{11}(t) &=& \rho_{00}(t_0) e^{- \Gamma_T(1 + 2\alpha) (t -t_0)
} +
\frac{\alpha}{1 + 2 \alpha}(1 - e^{- \Gamma_T(1 + 2\alpha) (t -t_0) }) \label{sol1}\\
 \rho_{01}(t) &=& \rho_{01}(t_0) e^{- i g B   (t - t_0) -
 (\Gamma_T (  \alpha + \frac{1}{2}) + \Gamma_L)(t - t_0)} \label{sol2}\\
 \rho_{00}(t) &=& 1 - \rho_{11}(t). \label{sol3}
\end{eqnarray}

If we include the translation degrees of freedom and consider an inhomogeneous magnetic field, Eq. (\ref{eq1}) generalizes to
\begin{eqnarray}
\frac{\partial \hat{\rho_1}}{\partial t} = - i g   B({\bf x})
[\hat{S}_z, \hat{\rho}_1] - \frac{\Gamma_T}{2} \alpha({\bf x})
\left([\hat{S}_-, [\hat{S}_+, \hat{\rho}_1]] + [\hat{S}_+
[\hat{S}_-,
\hat{\rho}_1]]\right) \nonumber \\
- \frac{\Gamma_T}{2}\left( [\hat{S}_-, \hat{\rho}_1 \hat{S}_+] -
[\hat{S}_+,
  \hat{S}_- \hat{\rho}_1]\right) -  \frac{\Gamma_L}{2} [\hat{S_z}, [\hat{S}_z,
\hat{\rho}_1]], \label{eq2}
\end{eqnarray}
where in this case $\hat{\rho}_1$ is a density matrix on the Hilbert space ${\cal H} = {\cal L}^2({\bf R}^3) \otimes {\bf C}^{2s+1}$.

In Eq. (\ref{eq2}) the parameter $\alpha$ depends on the position ${\bf x}$ due to its dependence on the inhomogeneous magnetic field
 $B$ as in Eq.
(\ref{alfa}). However, the $\alpha$ term does not change the phases generated during time evolution, which are the important variables in the spin-echo experiment. Hence, assuming that the field inhomogeneities are small, we may treat $\alpha$ as a constant.

\subsection{Time-evolution of entropy.}
From Eq. (\ref{eq2}) we construct the distribution
$f({\bf x}, {\bf p}, {\bf s})$ that describes the macrostates
according to Eq. (\ref{fq}). The corresponding Boltzmann entropy Eq.
 (\ref{We})   is expressed in terms of
the Wehrl entropy.
In order to compute the latter for  spin $s = \frac{1}{2}$, we
exploit the fact that the Wehrl entropy  is invariant
under $SU(2)$ transformations \cite{Wehrl}. An $SU(2)$
transformation can bring any density matrix $\hat{\rho}$ to its
diagonal form $\hat{\rho} = \mbox{diag} \left( x, 1-x \right)$,
parameterized by the single mixing parameter $ x \in [0, 1]$. Then,

\begin{eqnarray}
S_W[\hat{\rho}] = \frac{1}{2x -1} \left[ (1 - x)^2 (\ln(1 - x) -
\frac{1}{2}) - x^2 (\ln x - \frac{1}{2}) \right]. \label{We2}
\end{eqnarray}

We now consider an evolution of the system according to Eq.
(\ref{eq2}) together with the operation of the two external pulses
that characterize the spin-echo experiments. Initially, the system
of $N$ dipoles occupies volume $V$ and it is in a state of thermal
equilibrium in the presence of the magnetic field ${\bf B}({\bf
x})$. The probability distribution   factorizes as $f({\bf x},
{\bf p}, {\bf s}) = \frac{N}{V} g({\bf p}) u_{th}(\theta, \phi)$, where
$g({\bf p})$ is the Maxwell distribution for the momenta normalized
to unity, and where
 $u_{th}$
corresponds to a thermal state for the spin degrees of freedom at
temperature $T = \beta^{-1}$
\begin{eqnarray}
u_{th}(\theta, \phi) = \frac{1}{1 + 2 \alpha} ( \alpha + \sin^2
\frac{\theta}{2}). \label{u0}
\end{eqnarray}

At $t = 0$, a $\frac{\pi}{2}$-pulse acts upon the system sending the
positive $z$-axis into the positive $x$-axis. This induces a change
$u_{th} \rightarrow u_0$, where
\begin{eqnarray}
u_0(\theta, \phi) = \frac{1}{2} \left( 1 + \frac{1}{1 + 2 \alpha}
\sin \theta cos \phi \right).
\end{eqnarray}
The system then evolves under Eq. (\ref{eq2}). The ${\bf
p}$-dependent component is unaffected, while $u_0$ evolves to
\begin{eqnarray}
u_t(\theta, \phi, {\bf x}) = \sin^2 \frac{\theta}{2} + \left[
\frac{1}{2} e^{- \Gamma_T (1 + 2\alpha)t} + \frac{1+ \alpha}{1 + 2
\alpha} (1 - e^{- \Gamma_T ( 1 + 2 \alpha)t}) \right] \cos \theta
\nonumber \\ + \frac{1}{2} \frac{1}{1 + 2 \alpha} e^{- (\Gamma_T (
\alpha + \frac{1}{2}) +\Gamma_L)t} \sin \theta \cos [\phi -
\omega_p({\bf x}) t], \label{ut}
\end{eqnarray}
where $\omega_p({\bf x}) = g \mu B({\bf x})$. At $t = \tau$, a $\pi$-pulse acts by inverting the spin's direction; then the system
evolves again under Eq. (\ref{eq2}). So for $t >  \tau$,
\begin{eqnarray}
u_t(\theta, \phi, {\bf x}) = \sin^2 \frac{\theta}{2} + \left[
\frac{1}{2} e^{-  \Gamma_T (1 + 2\alpha) t } + \frac{1+ \alpha}{1
+ 2 \alpha} (1 - e^{-  \Gamma_T ( 1 + 2 \alpha) t}) \right] \cos
\theta
\nonumber \\
+    \frac{1}{2(1 + 2 \alpha)} e^{-  [\Gamma_T ( \alpha +
\frac{1}{2}) + \Gamma_L] t}\sin \theta \cos [\phi - \omega_p({\bf
x}) (t - 2 \tau)]. \label{ut2}
\end{eqnarray}

If  $\Gamma_T > 0$, Eq. (\ref{eq2}) does not preserve the energy, since the spin-lattice coupling transfers energy from the spin to the translational degrees of freedom. First, we consider the energy-preserving case $\Gamma_T = 0$, and we calculate the evolution of the Boltzmann entropy $S_B$ Eq. (\ref{entropy}). The dependence of $S_B$ on
$\omega_p({\bf x})$ cancels out due to the invariance of  the Wehrl entropy  under $SU(2)$ transformations. This
property is not affected by the position-dependence of the
transformations. We differentiate Eq. (\ref{entropy}) with respect to time and we obtain

\begin{eqnarray}
\dot{S}_B = \frac{N}{2(2\alpha+1)} e^{-\Gamma_Lt} \int \frac{d
\theta d \phi}{2 \pi} \sin^2 \theta  \cos \phi (1 + \ln u ) \geq 0.
\end{eqnarray}
The last step follows from the fact that the terms multiplying  the positive values of $\cos \phi$ are always larger than the terms multiplying negative values of $\cos \phi$ (
 $  u (\theta, \phi) \geq u(\theta, \pi - \phi)$, for $0\leq \phi \leq \pi$).
Therefore, we conclude that  the Boltzmann entropy  is an increasing function of time, and
the evolution is genuinely irreversible.

Next, we consider the general case $\Gamma_T > 0 $. We assume that the system is in contact with
 thermal reservoir of
temperature $\beta^{-1}$, coupling to the translational degrees of freedom. We also  assume that the relaxation time of the translational degrees of freedom is
 much shorter than $\Gamma_L^{-1}$. Hence, they always remain in a thermal state.
The energy $\Delta E$ lost by the spin degrees of
freedom is transferred to the thermal reservoir, and the
entropy of the reservoir increases by an amount $\Delta S_r \geq \beta
\Delta E$. The  total entropy change $\Delta S_{tot}$ satisfies the inequality
\begin{eqnarray}
\Delta S_{tot} \geq \Delta S_B + \beta \Delta E = \Delta (S_B -
\beta \bar{H}),
\end{eqnarray}
 where $S_B$ is the Boltzmann entropy and $\bar{H}$
is the mean value of the system's Hamiltonian. From Eq. (\ref{ut})
we find
\begin{eqnarray}
\bar{H}(t) = -  \frac{\bar{\omega}_p}{1 + 2 \alpha} \left( 1 - e^{-
\Gamma_T (1 + 2\alpha) t} \right)   ,
\end{eqnarray}
where $\bar{\omega}_p = \frac{1}{V} \int_V \omega_p(x) = g \mu
\bar{B}$ is the spatial average of the precession frequency.

Using Eqs. (\ref{We}, \ref{We2}), we find that $
S_B - \beta \bar{H}$ is an increasing function of $t$ for all values
of relaxation time and temperature. The behavior of $
S_B - \beta \bar{H}$ is plotted in Fig.
1 for different values of temperature. Consequently
\begin{eqnarray}
\frac{dS_{tot}}{dt} \geq \frac{d (S_B - \beta \bar{H})}{dt} \geq
0.
\end{eqnarray}
{\em  Hence, the total entropy  increases in time
during the spin echo experiment: there is no anti-thermodynamic
behavior and no violation of the second law of thermodynamics.}

\begin{figure}[tbp]
\includegraphics[height=6cm]{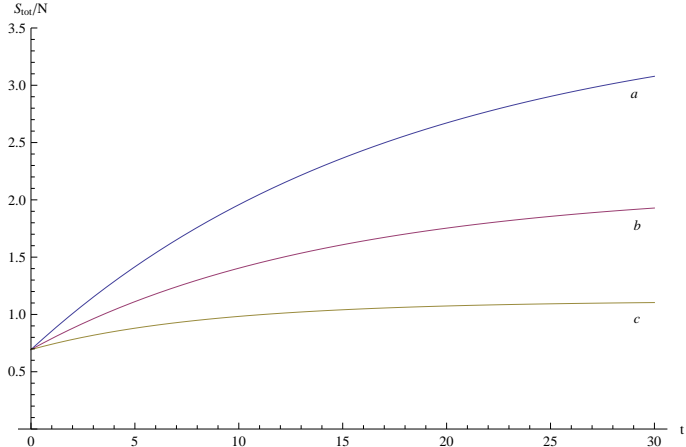} \caption{ \small (Color online) The minimum value of total entropy $S_{tot} = S_B - \beta \bar{H}$ plotted as an increasing function of time $t$, for $\Gamma_T/(g\mu B) = 0.05$ and different values of temperature $T$. Curve (a) corresponds to $T = 0.3g \mu B$, curve (b) to $T=0.5 g \mu B $ and curve (c) to $T = g \mu B$. }
\end{figure}

The above results are valid for the Markovian master equation
Eq. (\ref{ut2}) that governs the evolution of the system. The predicted
decay of the echo at time $t = 2 \tau$, defined as
\begin{eqnarray}
D(\tau) := \frac{M_x(2 \tau)}{M_x(0)} =   e^{- 2 [\Gamma_T ( 2\alpha + 1) +
\Gamma_L] \tau} = e^{-2\tau/T_2}, \label{dt}
\end{eqnarray}
is an exponential function.

The distribution Eq. (\ref{ut}) at time $t = 2 \tau$ depends on $\tau$ only through
exponentials. Hence, we perform a change of variables and we express the increase $\Delta S_{tot}(\tau)$  of the
total entropy  as a
function of the decay $D(\tau)$.
$\Delta S_{tot}$ is a strictly decreasing function of $D$---see Fig.
2. Hence, the common conjecture that the echo decay parameter provides a faithful measure of
entropy increase---for example, Refs. \cite{Prig, PL}---is verified.

\begin{figure}[tbp2]
\includegraphics[height=6cm]{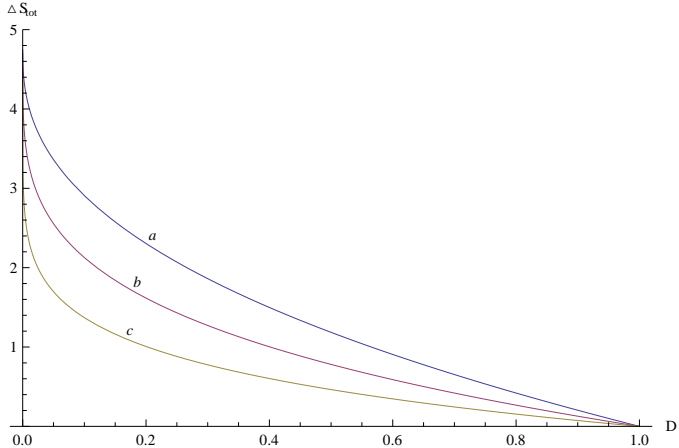} \caption{The  increase in the total entropy $\Delta S_{tot}$ during  an echo is a monotonously decreasing function
of the magnetization decay $D(\tau) = \frac{M_x(2 \tau)}{M_x(0)}$. Here, curve (a) corresponds to $T_1/T_2 = 0.75$, curve (b) to $T_1/T_2 =1.5$ and curve (c) to $T_1/T_2 = 3$. }
\end{figure}

\subsection{Loscmhidt echoes}

Here we considered  Hahn spin echoes, where the external pulses invert only the evolution by the free spin Hamiltonian---they do not affect the spin-spin and spin-lattice interactions. More complex pulse sequences may also achieve the inversion of spin-spin couplings, whence a more complete realization of Loschmisdt's idea is achieved \cite{Pines}. Moreover, echoes for localized excitations have been obtained. In general, inversion is never perfect, and the echo signal decays with  time. We emphasize here the existence of systems with fast spin dynamics, where the decay is not determined by any external interactions (e.g., spin-lattice coupling) but by the {\em reversible} spin dynamics itself \cite{PL} (Loschmidt echoes). In this case, the function $D(\tau)$ is best described by  Gaussian rather than exponential decay as in Eq. (\ref{dt}).

In this work, emphasis is given on the construction of a non-equilibrium entropy, according to Boltzmann, that provides a consistent thermodynamic description of spin-echo experiments. We showed that this entropy is a monotonic function of the echo decay $D(\tau)$.  We  defined the entropy by adapting Boltzmann's coarse-graining for the rare gases into the spin context: macrostates correspond to a distribution function on the state space of a single particle. There seems  no difficulty in applying the same construction to all echo experiments. In fact, we can take the analogy with Boltzmann's theory of rare gases one step further. We can construct an evolution equation for the macrostates, in analogy to Boltzmann's equation, where the spin-spin interaction is incorporated in   {\em non-linear} collision terms. The possibility that such an evolution equation  could account for the Gaussian decay law in Loschmidt-echo experiments is at present being explored.

\section{Discussion}

An important conclusion is drawn from the work presented here:  the interpretation of the
spin-echo experiments depends closely on the choice of macrostates
 one adopts for the system. In particular, if the
macroscopic description is at the level of the spin degrees of
freedom only, then the conclusion that the spin echo manifests an
anti-thermodynamic behavior is inevitable.
If, however, the macrostates refer also to the translational degrees
of freedom,  then the   loss of information due to dephasing   is compensated by non-trivial
correlations between the spin and position variables. If the chosen
macrostates   accommodate such correlations, then the corresponding
entropy does not exhibit a decreasing phase. When
dissipation effects are ignored, the  entropy of the system remains unchanged during time
evolution. The evolution law for the macrostates is reversible and the
dephasing is generated by volume-preserving Hamiltonian dynamics.

The main conclusion is strengthened by the consideration of
relaxation effects.  The
usual semi-phenomenological Bloch equations, describing relaxation in
magnetic systems, correspond to a description of irreversible
dynamics in terms of a Lindblad master equation. We showed
that the total entropy strictly increases as a function of time
under this dynamics. In effect, there is nothing extraordinary in
the thermodynamic behavior of spin echoes: Hamiltonian effects such
as  dephasing do not change the entropy and the relaxation effects
increase the entropy. There is no phase of decreasing entropy.

Our results have some interesting implications in relation to the foundations of statistical mechanics.
The properties of equilibrium thermodynamics are not
always reliable guides for the description of non-equilibrium
processes. In one sense, the apparent anti-thermodynamic behavior in
the spin-echo experiments is due to the fact that the concept of
entropy in equilibrium configurations is naively transferred into a
non-equilibrium context \cite{Lands}.   In an equilibrium spin system, the presence of a net transverse magnetization is
highly `improbable', hence it can be viewed as a witness of a low-entropy state. 
If such a state were considered to arise `spontaneously', one would then say that we have a
violation of the second law of thermodynamics. However, in the
non-equilibrium context there is no one-to-one correspondence between magnetization and entropy.

Our results also suggest  that macrostates are
not subjective: they do not correspond to a
description of the system in terms of variables that are {\em
accessible to us} through measurements of macroscopic variables.
In spin-echo experiments, the spin-position correlations are not
{\em directly} accessible, and one is, therefore, tempted to ignore
them in the treatment of the system. However, this is  an
approximation. It turns out that it is a drastic one: it ignores the correlations
between spin and position degrees of freedom and,
consequently, it misrepresents the thermodynamic behavior of the system.

We must emphasize that our definition of the macrostates in the spin-echo system is not arbitrary. Boltzmann's coarse-graining in terms of the state space of a single particle (with spin) can be immediately generalized to other set-ups, including relativistic systems. It is a natural definition, in the sense that the macrostates carry a representation of the fundamental symmetries of the system, and they arguably correspond to the external operations that can be effected on its constituents.


\begin{thebibliography}{}

\bibitem{Hahn} E. L. Hahn,  Phys. Rev. 80, 580 (1950).

\bibitem{Roth} J. Rothstein, Am. J. Phys. 25, 510 (1957).

\bibitem{Blatt} J. M. Blatt, Prog. Th.
Phys. 22, 745 (1959).


\bibitem{Prig} I. Prigogine, in {\em A Critical Review of Thermodynamics}, edited by E. Stuart, B. Gal-Or, and A.Brainard (Mono Books, Baltimore, 1970); I. Prigogine, C. George, F. Henin and L Rosenfeld, Chem.
Scr. 4, 5 (1973).

\bibitem{BH} R. G. Brewer and E. L. Hahn, Sc. Am. 251, 50 (1984).

\bibitem{DD85} K. G. Denbigh and J. S.  Denbigh, {\em Entropy in Relation to Incomplete
Knowledge} (Cambridge University Press, Cambridge, 1985).


\bibitem{SK}  L. Sklar, {\em Physics and Chance} (Cambridge University Press,
Cambridge, 1993).

\bibitem{Lands} P. T. Landsberg, Dialectica 50,   247 (1996).


\bibitem{RR} T. M. Ridderbos and M. L. G. Redhead, Found. Phys. 28,
1237 (1998); P. Ainsworth, Found. Phys. Lett. 18, 621 (2005).

\bibitem{Lav} D. A. Lavis, Found. Phys. 34, 669 (2004).


\bibitem{ZL} S. Loyd and W. H. Z. Zurek, J. Stat. Phys. 62, 819
(1991); K. Shizume, J. Stat. Phys. 70, 1572 (1993).


\bibitem{Losch} J. Loschmidt,   Sitzungsber. Kais. Akad.
Wiss. Wien Math. Naturwiss.  73,   128  (1876).


\bibitem{Pines} W. K. Rhim, A. Pines, and J. S. Waugh, Phys. Rev. B3, 684
(1971); J. S. Waugh, W. K. Rhim and A. Pines, J. Magn. Reson. 6, 317
(1972); J. S. Waugh, in {\em Pulsed magnetic resonance--NMR, ESR,
and optics: a recognition of E.L. Hahn} (Oxford University Press,
Oxford, 1992).

\bibitem{SZ} V.A. Skrevbnev and R.N. Zaripov, Appl. Magn. Reson. 16, 1 (1999).

\bibitem{PL}G. Usaj, H.M. Pastawski, P.R. Levstein, Mol. Phys. 95, 1229 (1998); P.R. Levstein, G. Usaj, H.M. Pastawski, J. Chem. Phys. 108, 2718 (1998);
H. M. Pastawski, R. P. Levstein, G. Usaj, J. Raya and J.
Hirschinger, Physica A283, 166 (2000).


\bibitem{Boltz} See, for example, R. S. de Groot and P. Mazur, {\em Non-Equilibrium
Thermodynamics}, (Dover, 1984); J. R. Dorfman, {\em An introduction
to chaos in nonequilibrium statistical mechanics} (Cambridge
University Press 1999). Also, J. L. Lebowitz, Physica A194, 1 (1993).


\bibitem{QBEW} D. Benedetto, F.
Castella, R. Esposito and M. Pulvirenti, J. Stat. Phys. 116, 381
(2004); 124, 951 (2005). For the Wigner function description of the
relativistic Boltzmann equation see, for example, E. Calzetta and B.
L. Hu, Phys. Rev. D 37, 2878  (1988).

\bibitem{QBE} N.M. Hugenholtz,   J. Stat. Phys. 32, 231 (1983); L. Erd\"os, M. Salmhofer
and H. T. Yau, J. Stat. Phys. 116, 367 (2004).


\bibitem{qpd} N. I. Balazs and B. K. Jennings, Phys. Rep. 104, 347 (1984); M. Hillery,
R. F. O'Connell, M. O. Scully, and E. P. Wigner, Physics Reports, 106, 121 (1984).

\bibitem{QMT} E. B. Davies, {\em Quantum theory of open systems}
(Academic Press, 1976); P. Busch, P. J. Lahti and P. Mittelstaedt,
{\em The Quantum Theory of Measurement}, (Springer, 1996).

\bibitem{CS} J. R. Klauder and B. S. Skagerstam, {\em Coherent States: Applications in Physics and Mathematical Physics} (
World Scientific, Singapore, 1985); A, M. Perelomov, {\em
Generalized Coherent States and Their Applications}, (Springer,
Berlin, 1986); W. M. Zhang, D. H. Feng and R. Gilmore, Rev. Mod.
Phys. 62, 867  (1990).

\bibitem {Omn8894} R. Omn\`es,   J. Stat. Phys.   53, 893, 1988;
\newblock  {\em The Interpretation of Quantum Mechanics},
\newblock  (Princeton University Press, Princeton, 1994);  Rev. Mod. Phys. 64, 339 (1992).


\bibitem{Wehrl} A. Wehrl, Rep. Math. Phys. 16, 353  (1979); E.H. Lieb,   Commun. Math.
Phys. 62, 35  (1978).

\bibitem{Hal}  J. J. Halliwell, Phys.Rev. D58,  105015 (1998);
 Phys. Rev. Lett. 83,  2481 (1999); Phys Rev D68, 025018 (2003).

\bibitem{Lind} G. Lindblad, Comm. Math. Phys. 48, 119 (1976).

\bibitem{Open} H. P. Breuer and F. Petruccione, {\em The theory of
quantum open systems}, (Oxford University Press, New York, 2002).



\end{thebibliography}
\end{document}